\documentclass[aps,prc,twocolumn,showpacs,floatfix]{revtex4-1}

\usepackage{graphicx}
\usepackage{dcolumn}
\usepackage{bm}
\usepackage{ulem}
\usepackage{color}

\newcommand{\al}{$\alpha$}
\newcommand{\g}{$\gamma$}

\newcommand{\rag}{($\alpha$,$\gamma$)}

\newcommand{\ovi}{$^{16}$O}

\newcommand{\nenull}{$^{20}$Ne}

\newcommand{\TZMC}{TZMC}

\begin{document}

\title{
  Comment on ``Observation of annual modulation by $\gamma$ rays from
  ($\alpha$,$\gamma$) reactions at the Soudan Underground Laboratory''
}

\author{Peter Mohr}
\email[Email: ]{mohr@atomki.mta.hu}
\affiliation{
Diakonie-Klinikum, D-74523 Schw\"abisch Hall, Germany}
\affiliation{
Institute for Nuclear Research (Atomki), H-4001 Debrecen, Hungary}

\date{\today}

\begin{abstract}
Tiwari {\it et al.}\ have identified an annual modulation of the $\gamma$-ray
flux at the Soudan Underground Laboratory which is strongly correlated to the
radon concentration. The $\gamma$-ray flux results from ($\alpha$,$\gamma$)
reactions which are induced by the $\alpha$ activity of radon and its
daughters. Unfortunately, the quantitative analysis of the $\gamma$-ray flux
is based on unrealistic ($\alpha$,$\gamma$) cross sections, and thus the
calculated $\gamma$-ray fluxes are not reliable.
\end{abstract}

\maketitle

In a recent study, Tiwari, Zhang, Mei, and Cushman (\TZMC )
\cite{Tiw17} have studied $\gamma$-ray fluxes which result from
\rag\ reactions on $^{16}$O, $^{27}$Al, and $^{28}$Si which are the most
abundant components of rock at the Soudan Underground Laboratory (SUL). The
\al\ particles are provided by \al\ decay of $^{220}$Rn and $^{222}$Rn and
all daughter nuclides (see Table I in \cite{Tiw17}). A correlation between the
annual modulation of the radon concentration at SUL and the \rag\ flux is
found by \TZMC . This is an important result which may affect the
interpretation of low-background experiments like the search for dark matter,
neutrionoless double-$\beta$ decay, or weakly interacting massive particles
\cite{Tiw17}. Whereas the temporal correlation between the radon
concentration and the \g -ray flux is well established, the quantitative
analysis suffers from unrealistic cross sections of the \rag\ reactions under
study by \TZMC .

The highest energy in the $^{220}$Rn chain is $E_\alpha = 8784$ keV
from the decay of $^{212}$Po; for $^{222}$Rn one finds $E_\alpha = 7687$ keV
from the decay of $^{214}$Po. (Note that Table I of \TZMC\ with the energies
$E_\alpha$ has been updated \cite{Tiw17}).

\TZMC\ use the TALYS code \cite{TALYS,TALYS2} to estimate the \rag\ capture
cross sections. The application of a statistical model code like TALYS for
light nuclei, in particular for the doubly-magic $^{16}$O with its low level
density, requires special care because the statistical model provides average
cross sections whereas in reality the \rag\ cross section is governed by
individual resonances. Limits of the applicability of the statistical model in
this mass range have recently been discussed in \cite{Mohr17}, and the isospin
suppression of \rag\ cross sections in $N=Z$ nuclei was analyzed in
\cite{Rau00b}. 

For simplicity, the following discussion mainly focuses on $^{16}$O and the
$^{16}$O\rag $^{20}$Ne reaction. The same arguments are also valid for
$^{27}$Al and $^{28}$Si.

As a first step, \TZMC\ use the TALYS code to calculate the cross section for
$\gamma$-ray emission at a given energy $E_\alpha = 4$ MeV (Fig.~5 in \TZMC
). A continuous $\gamma$-ray spectrum is found with energies from about 1 MeV
up to more than 14 MeV. Obviously, two conspiciuties result from the
application of the TALYS code and the chosen parameters: ($i$) The maximum
$\gamma$-ray energy in the $^{16}$O\rag $^{20}$Ne reaction is limited to
$E_{\gamma,{\rm{max}}} = E_{\rm{c.m.}} + Q = (16/20) E_\alpha + 4730$ keV =
7930 keV for $E_\alpha = 4$ MeV. This strict physical limit is at least 6 MeV
below the highest \g -ray energies shown in Fig.~5 of \TZMC . ($ii$) The
$\gamma$-ray spectrum in Fig.~5 should be discrete with primary lines at 7930
keV for ($\alpha$,$\gamma_0$), 6296 keV for ($\alpha$,$\gamma_1$) to the $2^+$
state at 1634 keV, 3682 keV for ($\alpha$,$\gamma_2$) to the $4^+$ state at
4248 keV, 2963 keV for ($\alpha$,$\gamma_3)$ to the $2^-$ state at 4967 keV,
etc. Secondary $\gamma$-rays appear at 1634 keV ($2^+ \rightarrow 0^+$), 2614
keV ($4^+ \rightarrow 2^+$), and 3333 keV ($2^- \rightarrow 2^+$); most of the
higher-lying states in $^{20}$Ne preferentially decay by \al -emission with
negligible secondary \g -rays.

For completeness, I also provide the maximum \g -ray energies for $E_\alpha =
4$ MeV in the $^{27}$Al\rag $^{31}$P and $^{28}$Si\rag $^{32}$S reactions
which are 13.15 MeV for $^{27}$Al and 10.45 MeV for $^{28}$Si. In both cases
\TZMC\ show \g -ray energies up to above 20 MeV in their Fig.~5.

From the discussion during the review process of this Comment, it became clear
that the \g -ray energies above $E_{\gamma,{\rm{max}}}$ in Fig.~5 of \TZMC\ 
result from the TALYS parameter ``{\it{elwidth}}'' which is the Gaussian
spreading width for outgoing particles. The intention of ``{\it{elwidth}}'' is
a simple comparison of the theoretical TALYS spectrum (with infinite energy
resolution) to experimental spectra (with finite energy resolution). By
default, ``{\it{elwidth}}'' is set to 0.5 MeV in TALYS, leading to a parabolic
high-energy tail in the logarithmic plot in Fig.~5 of \TZMC .

Next, \TZMC\ use the calculated \g -ray spectra at different energies
$E_\alpha$ to calculate the \g -ray yield according to their Eq.~(1). In the
case of $^{16}$O, this leads to Fig.~8 of \TZMC\ with a continuous \g -ray
spectrum up to $E_{\gamma,{\rm{max}}} \approx 18$ MeV. Again, the highest
energies can only be reached because of the above discussed ``{\it{elwidth}}''
parameter, as the highest $E_\alpha = 8784$ keV from the
$^{212}$Po decay leads to $E_{\gamma,{\rm{max}}} = 11757$ keV. Note that also
for the low-abundant $^{17,18}$O with their larger $Q$-values
$E_{\gamma,{\rm{max}}}$ remains below 16854 keV.

\begin{figure}[thb]
\includegraphics[bbllx=30,bblly=31,bburx=400,bbury=756,width=0.85\columnwidth,clip=]{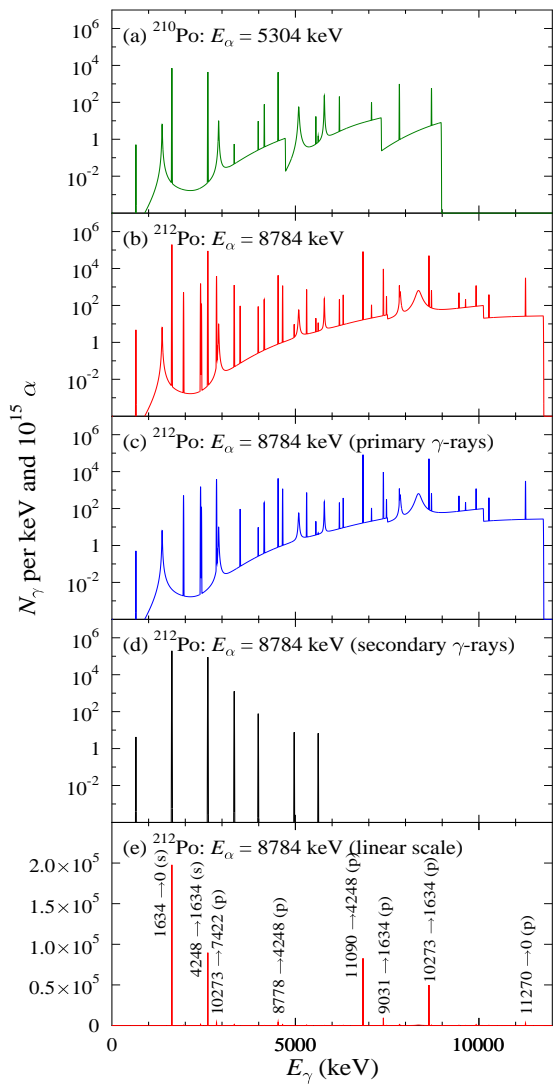}
\caption{
\label{fig:spec_o16ag}
(Color online) 
Thick target \g -ray spectra of \ovi \rag \nenull\ for $E_{\alpha,{\rm{lab}}}
= 5304$ keV (a) and 8784 keV (b-e), corresponding to the \al -decays of
$^{210}$Po and $^{212}$Po. At the higher energy of 8784 keV the total spectrum
(b) is decomposed into the primary (c) and secondary (d) \g -rays; in
addition, the total \g -ray spectrum is shown in linear scale (e) with the
dominating lines labeled by the level energies (in keV) and (p) or (s) for
primary or secondary \g -rays.
}
\end{figure}
For a smooth excitation function of the $^{16}$O\rag $^{20}$Ne reaction, the
energy loss of the \al\ particles in the rock at SUL would indeed finally lead
to a continuous \g -ray spectrum for the primary \g -rays. (The secondary \g
-rays appear as discrete lines in any case and are missing in Fig.~8 of \TZMC
). However, the excitation function of the $^{16}$O\rag $^{20}$Ne reaction is
not smooth, but governed by individual resonances (see e.g.\ Figs.~3 and 4 of
\cite{Pea64} in the relevant energy range of the present study) which are
superimposed over a small non-resonant contribution (e.g.,
\cite{Cos10,Hag11,Hag12,Mohr05}). Thus, \g -rays with discrete energies from
about 15 resonances with their individual branchings should be visible in the
\g -ray flux in Fig.~8 of \TZMC . This holds even for the heavier targets
under study by \TZMC , as can e.g.\ be seen in the \g -ray spectrum of a
thick-target experiment of the $^{28}$Si\rag $^{32}$S reaction at $E_\alpha
\approx 3$ MeV (see Fig.~1 of \cite{Bab02}).

As an illustration of the expected discrete $\gamma$-ray spectra,
Fig.~\ref{fig:spec_o16ag} shows the calculated thick-target $\gamma$-ray
yields from the \ovi \rag \nenull\ reaction for the energies
$E_{\alpha,{\rm{lab}}} = 5304$ keV and 8784 keV, corresponding to the \al
-decay energies of $^{210}$Po and $^{212}$Po; these are the highest and lowest
energies under study in \TZMC .  The calculated spectra in
Fig.~\ref{fig:spec_o16ag} are based on experimental resonance properties (as
adopted in \cite{Til98}) and complemented by a small non-resonant direct
capture contribution (taken from the calculations in \cite{Mohr05}, in good
agreement with experiment at $E_{\rm{c.m.}} = 2.345$ MeV \cite{Hag12}). The \g
-ray yield is dominated by very few strong resonances at higher energies and
two secondary \g -ray lines below 4 MeV; this becomes nicely visible in the
linear plot (e) of Fig.~\ref{fig:spec_o16ag}. All \g -ray spectra in
Fig.~\ref{fig:spec_o16ag} are shown for a resolution of 5 keV.

Similar spectra for the $^{27}$Al\rag $^{31}$P and $^{28}$Si\rag $^{32}$S
reactions cannot be calculated because the resonance properties are not
completely known for the relevant energy range. However, a measurement of
these spectra seems feasible: the \g -ray spectra for the \ovi \rag
\nenull\ reaction in Fig.~\ref{fig:spec_o16ag} are shown for $10^{15}$ \al
-particles which corresponds to a few minutes of a 1 $\mu$A \al\ beam.

Finally, it should be noted that Eq.~(1) of \TZMC\ differs from the
calculations in Heaton {\it et al.}\ \cite{Hea89} (Ref.~$[19]$ in
\TZMC ): As usual, in \cite{Hea89} (p.~534) the yield results from an integral
over $\sigma(E)/S(E)$ (with the energy-dependent stopping power $S(E)$ in the
integrand) whereas Eq.~(1) of \TZMC\ only contains the stopping power at the
initial energy ($E_j$ in \TZMC ) outside the integral.

At the very end, \TZMC\ provide integrated \g -ray fluxes in the broad energy
interval from 4 to 10 MeV (Table II in \cite{Tiw17}). This energy interval
roughly excludes their non-physical results where the calculated $E_\gamma$ is
above the maximum energy $E_{\gamma,{\rm{max}}} = E_{\rm{c.m.}} + Q$, and it
integrates over the discrete resonances which should have appeared in
Figs.~$6-8$ of \TZMC . Most secondary \g -ray lines show lower energies than 4
MeV and do not affect the $4-10$ MeV energy window of \TZMC . Thus, in
conclusion I point out that the final \g -ray fluxes of \TZMC\ may have the
correct order of magnitude although the quantitative analysis of \TZMC\ is
inconsistent. Because of the importance of the results for the understanding
of the background at SUL, an improved analysis should be performed.

\noindent
This work was supported by NKFIH (K108459 and K120666).

\end{document}